\documentclass[final,5p,times,twocolumn]{elsarticle}
\usepackage{amssymb}
\usepackage{lipsum}
\usepackage{natbib}
\usepackage{caption}
\usepackage{siunitx}
\usepackage[pdfborder={0 0 0},colorlinks, linkcolor=blue, citecolor=green]{hyperref}
\usepackage{chemformula}
\journal{Nuclear Instruments and Methods in Physics Research}
\begin{document}
\begin{frontmatter}
%% Title, authors and addresses

%% use the tnoteref command within \title for footnotes;
%% use the tnotetext command for theassociated footnote;
%% use the fnref command within \author or \affiliation for footnotes;
%% use the fntext command for theassociated footnote;
%% use the corref command within \author for corresponding author footnotes;
%% use the cortext command for theassociated footnote;
%% use the ead command for the email address,
%% and the form \ead[url] for the home page:
%% \title{Title\tnoteref{label1}}
%% \tnotetext[label1]{}
%% \author{Name\corref{cor1}\fnref{label2}}
%% \ead{email address}
%% \ead[url]{home page}
%% \fntext[label2]{}
%% \cortext[cor1]{}
%% \affiliation{organization={},
%%            addressline={}, 
%%            city={},
%%            postcode={}, 
%%            state={},
%%            country={}}
%% \fntext[label3]{}

\title{Integrating Superregenerative Principles in a Compact, Power-Efficient NMR/NQR Spectrometer: A Novel Approach with Pulsed Excitation}
% authors are listed alphabetically Thorsten will decide before presenting this to Andrzej
\author{Tomas Sikorsky}
\author{Andrzej Pelczar}
\author{Stephan Schneider}
\author{Thorsten Schumm}
\affiliation[first]{organization={Atominstitut, TU Wien},addressline={Stadionallee 2}, 
            city={Vienna},
            postcode={1020},
            country={Austria}}

\begin{abstract}
We present a new approach to Nuclear Quadrupole Resonance (NQR)/Nuclear Magnetic Resonance (NMR) spectroscopy, the Damp-Enhanced Superregenerative Nuclear Spin Analyser (DESSA). This system integrates Superregenerative principles with pulsed sample excitation and detection, offering significant advancements over traditional Super-Regenerative Receivers (SRRs). Our approach overcomes certain limitations associated with traditional Super-Regenerative Receivers (SRRs) by integrating direct digital processing of the oscillator response delay time (T$_d$) and an electronic damp unit to regulate the excitation pulse decay time (T$_e$). The essence is combining pulsed excitation with a reception inspired by, but distinct from, conventional SRRs. The damp unit allows a rapid termination of the oscillation pulse and the initiation of detection within microseconds, and direct digital processing avoids the need for a second lower frequency which is used for quenching in a traditional SRRs, thereby avoiding the formation of sidebands. We demonstrate the effectiveness of DESSA on a \ch{NaClO3} sample containing the isotope Chlorine-35 where it accurately detects the NQR signal with sub-kHz resolution.
\end{abstract}

%%Graphical abstract
%\begin{graphicalabstract}
%\includegraphics{grabs}
%\end{graphicalabstract}

%%Research highlights
%\begin{highlights}
%\item Research highlight 1
%\item Research highlight 2
%\end{highlights}

\end{frontmatter}
\section{Introduction}
Efforts to develop more efficient and practical detection techniques are at the forefront of research in Nuclear Quadrupole Resonance (NQR) and Nuclear Magnetic Resonance (NMR) spectroscopy~\citep{sugiki2018current,schultz1971applications,Gilby1970}. NMR/NQR were initially developed as a continuous wave (CW) methods, using a static magnetic field and varying the frequency of the RF probe field~\citep{purcell1946resonance,pound1950nuclear}. However, by the 1950s, the CW methods were superseded mainly by pulsed techniques that offered higher resolution and greater sensitivity~\citep{hahn1950spin}. These methods, which include Fourier-transform NMR, made use of short pulses of the RF excitation field and measurement of the resulting Free Induction Decay (FID) signals, providing a greater wealth of information about the sample~\cite{ernst1966application}. In the 1980s, a method called NQR began to gain recognition for its potential in detecting explosives and narcotics. It is conceptually analogous to NMR but detects the interaction of the nuclear quadrupole moment with the sample-intrinsic electric field gradient at the nucleus position, in contrast to nuclear dipole moments interacting with externally applied magnetic field in NMR~\citep{garroway2001remote,latosinnska2007applications}.

The Super-Regenerative Receiver (SRR) is a detection method that was originally used in Continuous Wave (CW) NQR spectrometers~\citep{armstrong1922some,dehmelt1950pure}. However, the SRR has limitations related to frequency stability~\citep{hughes1975new}, a potential for generating sidebands of detected NMR/NQR signals~\citep{rehn1960side}, and line shape distortions~\citep{moncunill2005generic,bohorquez2009frequency,smith1968improved}. These drawbacks have limited the performance of the SRR in various applications~\citep{melnick1980determination}. Our work focuses on enhancing the traditional SRR method by integrating it with pulsed excitation. 

Traditional super-regenerative receiver (SRR) spectrometer operation is based on regenerative amplification. Absorption of radiation by the sample affects the rate of decay of oscillations and therefore the level remaining at the moment of successive pulse buildup~\citep{graybeal1958nuclear}. This affects the behavior of the circuit, which is periodically interrupted by a quench signal~\citep{hoffmann1979pocket}. The quenching creates a series of decaying oscillations whose amplitude is proportional to the strength of the received signal~\citep{bogaard1964quench}. This process leads to amplification, allowing for the detection of NMR/NQR signals~\citep{narath1964low}.

Our method differs conceptually from this operation mode. The conventional SRR methodology involves continuous amplification and quenching on some sampling frequency that requires subsequent demodulation. Our technique directly measures the delay time (T$_d$), which is the time interval between the activation of the quench (interrupting the sample excitation) and the onset of the subsequent oscillation triggered by the signal emitted by the sample (see \autoref{fig:timing}). In doing so, we not only avoid the need for demodulation, which prevents the formation of sidebands, but also ensure that signal detection depends solely on the induced response of the sample, rather than absorption. Similar to pulsed NMR techniques, our method uses a transmitted excitation pulse. For the response reception, we use SRR principles to measure the amplitude of the NMR/NQR signal at a very short interval around point t$_0$ (as shown in \autoref{fig:timing}). In SRR, the sensitivity period is the short time interval around the closure of the quench unit and the reactivation of the oscillatory unit (t$_0$). The delay time (T$_d$) of the pulse is mainly influenced by the signal amplitude during the sensitive period~\citep{moncunill2005generic,hernandez2019analysis}. Our design is enhanced and automated by direct digital processing of the regeneration time delay (T$_d$) and the use of a dedicated electronic damping circuit to shorten the decay time (T$_e$) of the excitation pulse.
\section{Design and Features}
We present the Damp-Enhanced Superregenerative Nuclear Spin Analyser (DESSA), a spectrometer designed for low power consumption, compactness and versatility. DESSA is capable of recording NMR/NQR spectra, revealing resonances and their linewidths from NMR/NQR active materials. The DESSA has most of its components on a single circuit board. The oscillator unit is generating the excitation RF signal. The damp and quench units are controlling the spectrometer's behavior. A transceiver coil mounted on the board both excites and detects the resonance signals of the sample. A control unit is galvanically isolated from the main board. This unit controls the damp and quench switches and contains a frequency adjustment module. All components are enclosed in a cooper Faraday cage, shielding the DESSA against external electromagnetic noise such as radio stations.
\begin{figure}[h!]
  \centering
  \includegraphics[width=\linewidth]{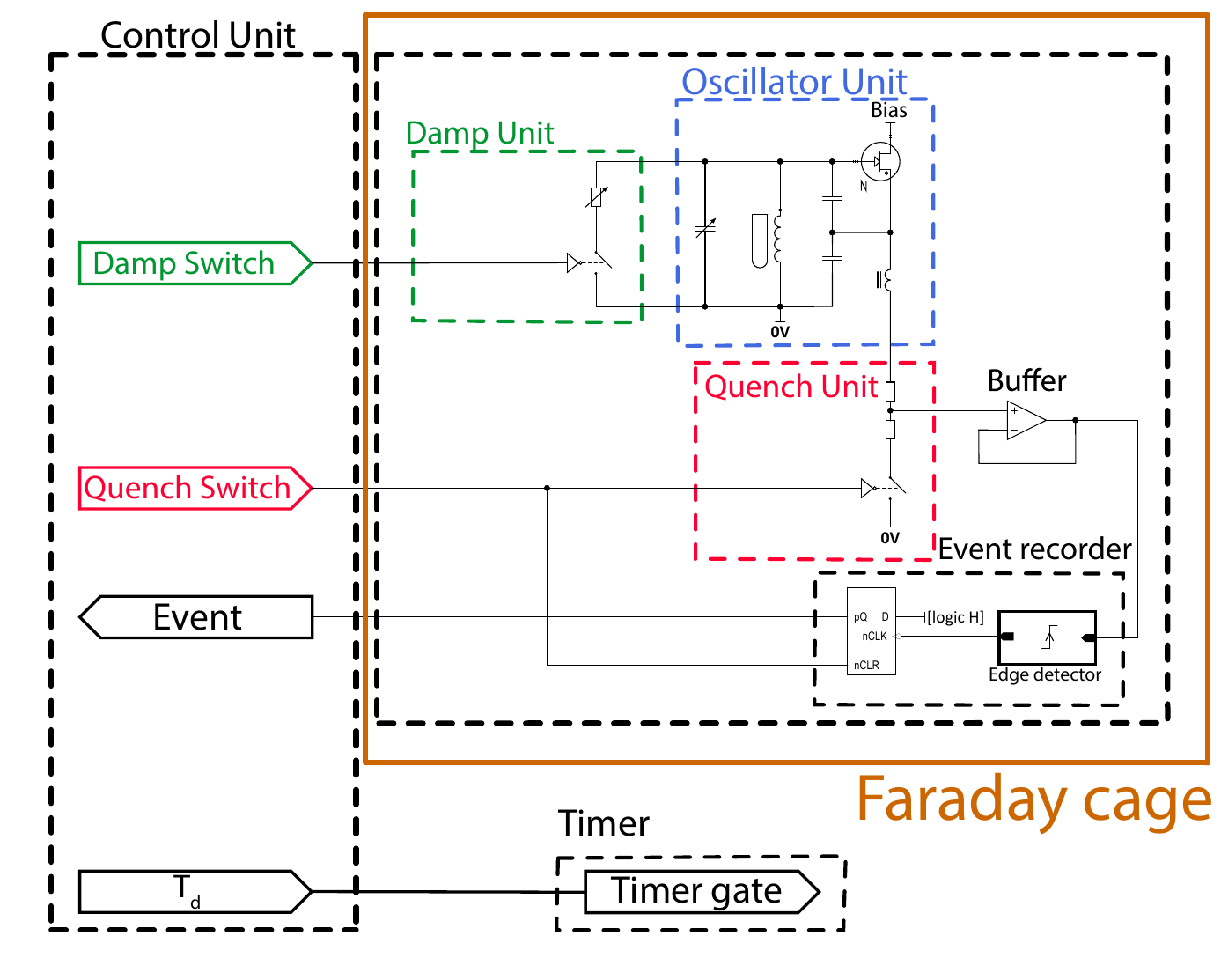} % adjust the file name as necessary
  \caption{Illustrates the layout of the DESSA units. The oscillator unit, situated close to the sample, is responsible for generating the excitation pulse. The event recorder, logs key events within the experimental sequence. The quench and damp switches, used for controlling the RF pulse, are represented in green and red, respectively. The control unit operating the oscillator, quench, and damp switches. All key components are mounted on a single circuit board and enclosed in a Faraday cage to prevent interference from external electromagnetic disturbances.}
  \label{fig:circuit} % label for cross-referencing
\end{figure}

The primary component of the DESSA is the oscillator unit. Upon activation, this oscillator produces an RF pulse to interact with the nuclear spins of the target material. When this RF pulse is resonant with the nuclear spin transitions in the sample, the spins are excited from their equilibrium state. After the end of the RF pulse, these spins generate an oscillatory response signal at their resonance frequency.

After the suppression of the RF oscillation pulse by the quench and damp units, the oscillator unit is re-activated, allowing for the formation of a subsequent oscillation pulse. If the on-resonance condition was satisfied, the response signal from the sample facilitates a faster buildup of the next oscillation pulse. The faster pulse buildup is an indicator of resonance match.
\begin{figure}[h!]
  \centering
  \includegraphics[width=\linewidth]{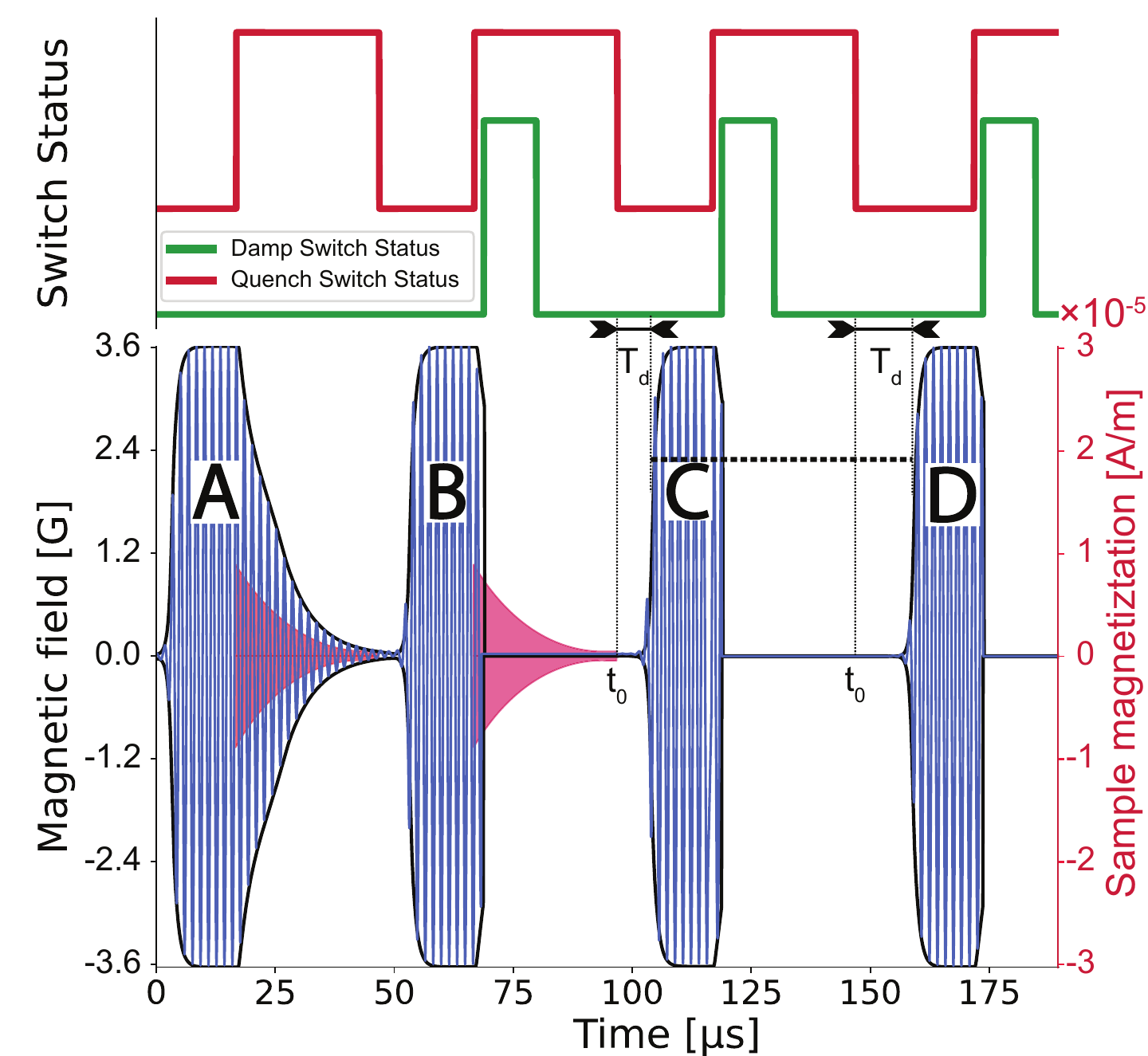} % adjust the file name as necessary
  \caption{Illustrates the behavior of the oscillator unit to quench and damp units. The upper part of the figure shows the status of the damp and quench switches with green and red lines, respectively. Below, the oscillatory behavior (blue) and envelope (black) illustrate the behavior in three distinct scenarios. Pulse A shows a standard quenching process without the damp unit, where the generated FID signal (red) is masked by a slowly decaying residual RF signal from the excitation pulse. Pulse B decays faster as a result of the damp unit, and the NMR/NQR signal amplitude is greater than the residual signal from the excitation pulse. The off-resonance pulse C leads to a longer delay time of the following pulse D due to an absence of a sample response. The short delay time of the pulse B is due to the absence of damping. The sensitive period of the receiver is marked as t$_0$. Oscillations are depicted with their frequency scaled down by a factor of 100 (blue curve). The FID, has a relaxation time of T2$^*$=10\,$\mu$s. The relaxation has been shortened for illustration purposes. The threshold that determines the rising edge of the oscillation buildup is show as black dashed line. The figure illustrates the behavior of DESSA's oscillator unit under different conditions. It does not represent an experimental sequence.}
  \label{fig:timing} % label for cross-referencing
\end{figure}

The analog signal edge detection element monitors the oscillator amplitude. This element activates when the amplitude surpasses a predefined voltage threshold. Its principal function is to determine (T$_d$), which is the interval between the moment the quench unit is deactivated and the instant the analog signal edge detection element acknowledges the amplitude surpassing the predefined threshold (see \autoref{fig:timing}). The direct digital processing of T$_d$ value indicates the degree of excitation of the target sample nuclei by a reduction of (T$_d$) in the case of a resonance.

Essential to the DESSA's design are the quench and damp units. The quench unit's role is to turn off the oscillator unit, which halts the energy flow into the oscillator. However, the energy in the circuit reverberates and decays on the timescale of tens of $\mu$s (see \autoref{fig:timing}). The damp unit, working in tandem with the quench unit, puts the circuit into a critically damped mode, which is the fastest method for damping the oscillations. This is essential because the response signal produced but a sample (if excited on resonance) is many orders of magnitude smaller in magnitude than the oscillation pulse, and the majority of NMR/NQR signals decay on the timescale of tens of $\mu$s due to fast spin-spin relaxation (T$_2^*$). By minimizing the time between the end of the excitation pulse and the beginning of the detection phase, we have significantly enhanced the sensitivity of DESSA.

The DESSA can be tuned to different frequencies by using a variable capacitor. This allows for precise tuning of the oscillator unit's resonance frequency, which is essential for detecting different nuclear spin resonances. This makes the DESSA an effective tool for various spectroscopic applications. The control unit is essential to its operation as a spectrometer. It records and stores the T$_d$ values as a function of the scanned frequencies. The data obtained is then compared with a frequency-dependent baseline that varies smoothly (mostly linearly) with frequency. When the T$_d$ value decreases below this baseline, it indicates a decrease in delay time, signifying the on-resonance excitation of the nuclear spin resonance with the DESSA's frequency.

\begin{figure}[h!]
  \centering
  \includegraphics[width=\linewidth]{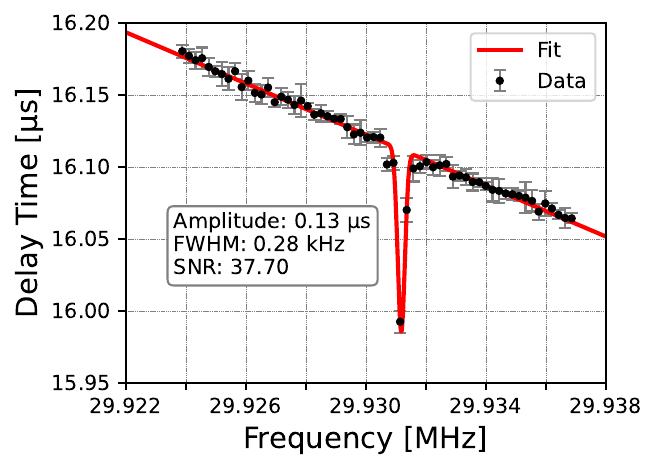} % adjust the file name as necessary
  \caption{The delay time response of a super-regenerative receiver (SRR) excited by varying frequencies, using \ch{NaClO3} as the sample. Each data point is an average of 2048 repetitions, with an excitation time of 160 microseconds and a recovery time of 100 milliseconds. The data was fitted with a Gaussian curve with a linear baseline.}
  \label{fig:response} % label for cross-referencing
\end{figure}

The DESSA NMR/NQR spectrometer is designed to operate in a small flip angle regime. This means that the nutation frequency is either smaller or comparable to the inverse of the sample's relaxation time (T$_2^*$), indicating a weak coupling. With a modest power consumption of only 0.5 watts for the oscillator unit and 5 watts for the control unit. The low power operation also allows for battery operation, making it an ideal device for applications in outdoor or low-energy environments.

To benchmark the sensitivity of the DESSA, we conducted a test using an anhydrous crystalline sodium chlorate (\ch{NaClO3}) sample. The sample weighed approximately 0.2\,grams and was housed in a standard 5\,mm NMR tube. The DESSA's oscillator unit was tuned to the nuclear quadrupole resonance frequency of the Chlorine-35, the transition from \(m = \frac{1}{2}\) to \(m = \frac{3}{2}\). The experiment was performed at ambient temperature of \(20\,^\circ C\). The nutation frequency was 1.5\,kHz and the effective flip angle was around 12$^{\circ}$. The cooper solenoid coil with 13 turns and 8\,mm length gave the Q factor of around 100. The resulting signal has an excellent signal-to-noise ratio and sub-kHz resolution. The resolution is limited by the sample temperature stability, rather than by the spectrometer. \ch{NaClO3} drifts at \SI{3}{\kilo\hertz\per\kelvin} at room temperature~\citep{wang1955pure}.

\begin{figure}[h!]
  \centering
  \includegraphics[width=\linewidth]{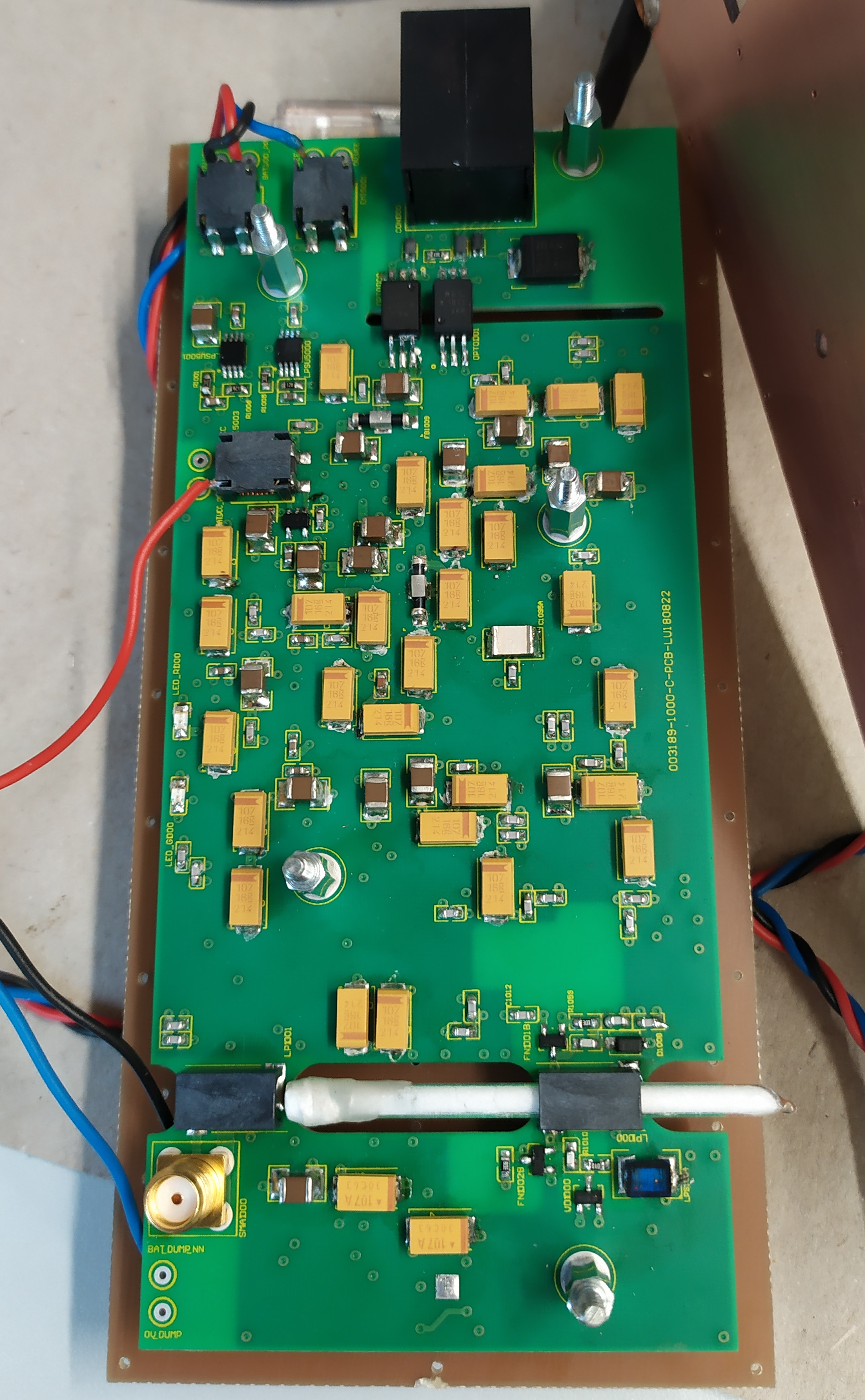} % adjust the file name as necessary
  \caption{The photo shows the DESSA circuit board with a white cylindrical sample tube mounted on copper blocks.}
  \label{fig:photo} % label for cross-referencing
\end{figure}

\section{Outlook and summary}
The DESSA brings a new way to detect nuclear spin resonance by combining established techniques with the advantages of compactness (measuring just one cubic decimeter) and low power consumption (averaging around 5.5 watts). DESSA has many potential applications, particularly in quality control, analytical chemistry, drug development, and magnetic resonance imaging. Its compactness and low power requirements make it an ideal solution for outdoor and in-field research. Due to its low RF emissions, the device is well suited for environments that are sensitive to radio frequency (RF) pulses.

\section*{Acknowledgements}
We thank Julia Pfitzer and Herman Scharfetter for the characterization of the \ch{NaClO3} NQR properties, and to René Sedmik and Michael Bartokos for enriching discussions. We would also like to express our gratitude to Martin Pressler and Jan Welsch for their work in preparing the \ch{NaClO3} sample. This work is part of the thorium nuclear clock project that has received funding from the European Research Council (ERC) under the European Union's Horizon 2020 research and innovation programme (Grant Agreement No. 856415). The research was supported by the Austrian Science Fund (FWF) Projects: I5971 (REThorIC) and P 33627 (NQRclock). We would like to acknowledge that the technology used in this study is subject to a patent pending status (Patent Application No. A50991/2023)

%% If you have bibdatabase file and want bibtex to generate the
%% bibitems, please use
%%
\bibliographystyle{elsarticle-num} 
\bibliography{main}

%% else use the following coding to input the bibitems directly in the
%% TeX file.

%\begin{thebibliography}{00}

%% \bibitem[Author(year)]{label}
%% For example:

%% \bibitem[Aladro et al.(2015)]{Aladro15} Aladro, R., Martín, S., Riquelme, D., et al. 2015, \aas, 579, A101

%%\end{thebibliography}

\end{document}